\documentclass[10pt, paper=a4, UKenglish]{article}
\usepackage{graphicx}
\def\Title#1{\begin{center} {\Large #1 } \end{center}}
\def\Author#1{\begin{center}{ \sc #1} \end{center}}
\def\Address#1{\begin{center}{ \it #1} \end{center}}

\newcommand\pubblock{\rightline{\begin{tabular}{l} Proceedings of the CTD/WIT 2019\\ \pubnumber\\
         \pubdate  \end{tabular}}}

\newenvironment{Abstract}{\begin{quotation} \begin{center} 
             \large ABSTRACT \end{center}\bigskip 
      \begin{center}\begin{large}}{\end{large}\end{center} \end{quotation}}

\newenvironment{Presented}{\begin{quotation} \begin{center} 
             PRESENTED AT\end{center}\bigskip 
      \begin{center}\begin{large}}{\end{large}\end{center} \end{quotation}}

\def\Acknowledgements{\bigskip  \bigskip \begin{center} \begin{large}
      \bf ACKNOWLEDGEMENTS \end{large}\end{center}}

\def\beq{\begin{equation}}
\def\eeq#1{\label{#1}\end{equation}}
\def\eeqn{\end{equation}}

\def\beqa{\begin{eqnarray}}
\def\eeqa#1{\label{#1}\end{eqnarray}}
\def\eeqan{\end{eqnarray}}

\let\bar=\overbar

\def\Dslash{\not{\hbox{\kern-4pt $D$}}}
\def\dslash{\not{\hbox{\kern-2pt $\del$}}}

\def\msb{{\bar{\ssstyle M \kern -1pt S}}}

\usepackage{color}
\usepackage{lineno}
\usepackage{subfig}
\usepackage{hyperref}

\newcommand\pubnumber{PROC-CTD19-008}

\newcommand\pubdate{\today}

\def\affiliation{
$^1$ Fermilab, Batavia, IL, USA 60510-5011

$^2$ Princeton University, Princeton, NJ, USA 08544

$^3$ University of Oregon, Eugene, OR, USA 97403

$^4$ University of California, San Diego, La Jolla, CA, USA 92093

$^5$ Cornell University, Ithaca, NY, USA 14853}

\newcommand{\conference}{Connecting the Dots and Workshop on Intelligent Trackers (CTD/WIT 2019)\\
Instituto de F\'isica Corpuscular (IFIC), Valencia, Spain\\ 
April 2-5, 2019}

\usepackage{fancyhdr}
\definecolor{mygrey}{RGB}{105,105,105}
\begin{document}


\large
\begin{titlepage}
\pubblock

\vfill
\Title{Speeding up Particle Track Reconstruction in the CMS Detector using a Vectorized and Parallelized Kalman Filter Algorithm}
\vfill

\Author{G. Cerati$^1$, P. Elmer$^2$, B. Gravelle$^3$, M. Kortelainen$^1$, V. Krutelyov$^4$, \\
S. Lantz$^5$, M Masciovecchio$^4$, K. McDermott$^5$, B. Norris$^3$, \\
M. Reid$^5$, A. Reinsvold Hall$^1$, D. Riley$^5$, M. Tadel$^4$, P. Wittich$^5$, \\
F. W\"{u}rthwein$^4$ and A. Yagil$^4$}
\Address{\affiliation}
\vfill

\begin{Abstract}
Building particle tracks is the most computationally intense step of event reconstruction at the LHC. With the increased instantaneous luminosity and associated increase in pileup expected from the High-Luminosity LHC, the computational challenge of track finding and fitting requires novel solutions. The current track reconstruction algorithms used at the LHC are based on Kalman filter methods that achieve good physics performance. By adapting the Kalman filter techniques for use on many-core SIMD architectures such as the Intel Xeon and Intel Xeon Phi and (to a limited degree) NVIDIA GPUs, we are able to obtain significant speedups and comparable physics performance. New optimizations, including a dedicated post-processing step to remove duplicate tracks, have improved the algorithm's performance even further. Here we report on the current structure and performance of the code and future plans for the algorithm.\end{Abstract}

\vfill

\begin{Presented}
\conference
\end{Presented}
\vfill
\end{titlepage}
\def\thefootnote{\fnsymbol{footnote}}
\setcounter{footnote}{0}
%

\normalsize 


\section{Introduction}
\label{intro}

As the high energy physics community prepares for the beginning of the High-Luminosity LHC (HL-LHC) data-taking period in 2026, one of the main questions that needs to be answered is how experiments such as CMS~\cite{CMS} will cope with the dramatic increase in computing requirements. In particular, the time it takes to reconstruct a single proton-proton bunch crossing (``event") is expected to go up by at least an order of magnitude due to the increased number of overlapping proton-proton collisions (pileup, PU). The average pileup will be approximately PU 50 in Run 3 of the LHC but will eventually increase to PU 200 during the HL-LHC.  

The largest contributor to the reconstruction CPU time is the charged particle track reconstruction, known as tracking, which has traditionally been performed using the Kalman filter method~\cite{KF}. At PU 50, tracking takes up nearly 60\% of the total reconstruction time for CMS events, and this fraction is expected to increase for larger values of pileup. Tracks are an essential input to many of the physics quantities of interest and are used to cluster jets, determine charged particle momenta, calculate missing transverse momentum, and tag heavy flavor jets. 

The mkFit group was formed in 2014 with the goal of writing a charged particle tracking algorithm that has a physics performance comparable to the current CMS tracking algorithm~\cite{CMS_tracking} but which is significantly faster. In order to accomplish this goal, the algorithm must take full advantage of highly parallel computing architectures such as the Intel Xeon Phi, Intel Xeon SP (Scalable Processors), and NVIDIA GPGPUs. A history of the mkFit project is given in Reference~\cite{Matevz}. This document highlights the latest developments and current performance of the algorithm on multicore CPUs. Work on a dedicated GPGPU implementation (see Reference~\cite{GPU} for previous results) is ongoing but will not be described in this document

\section{Parallel Kalman filter tracking}
\label{mkFit}

The mkFit algorithm is a parallelized and vectorized implementation of the Kalman filter method~\cite{KF}. The parallelization is performed using the Threading Building Blocks (TBB) library from Intel. Parallelization occurs at multiple levels. First, several events can be processed simultaneously. Second, within each event the seed tracks are divided into five pseudorapidity regions. Finally, within each region, the seed tracks are processed in batches of 16 or 32 seeds per batch.

The vectorization of the algorithm is done via a custom library called \textsc{Matriplex}. Unlike most high performance computing applications, the computations done by the Kalman filter algorithm are primarily matrix operations of very small matrices. The \textit{largest} matrices used in the Kalman filter tracking algorithm are symmetric 6$\times$6 matrices representing the covariance matrix of the track parameters. The \textsc{Matriplex} library was designed to efficiently vectorize these matrix operations. It uses a ``matrix-major" representation, where the first element of $N$ matrices are placed in the vector register and processed simultaneously. It can be used to generate either C++ code or intrinsics for multiplication of matrices of a given dimension. Additionally, \textsc{Matriplex} has the ability to be told in advance about known 0 or 1 elements in a matrix, which reduces the required number of operations by up to 40\%. 

\section{Application to the CMS detector}
\label{cms}

The goal for the mkFit algorithm is for it to be integrated into CMSSW, the software framework of the CMS experiment, and employed in the High Level Trigger (HLT), and possibly in offline reconstruction, during the HL-LHC. There are two setups that can be used to run the mkFit algorithm: either standalone or integrated with CMSSW. The standalone setup is used for development and for validation of the compute and physics performance of the algorithm. The integrated setup is used to directly compare the performance of the mkFit algorithm to that of the nominal CMSSW algorithm. When integrated, mkFit is run as an external software package. A dedicated CMSSW processing module converts the input hit and seed data into the format required by mkFit before track building and converts the tracks back into the CMSSW format after track building. The short term goal is to optimize this data conversion to speed up the mkFit algorithm when run within CMSSW. The longer term goal is to coordinate with the upstream and downstream algorithms in CMSSW to agree on a common data format in order to remove the need for data conversion entirely.

The results in this document use the Phase-I CMS detector geometry. The geometry is implemented as a plugin, independent of the core Kalman filter algorithm. Defining the geometry includes specifying the hit search windows, physical dimensions of a layer, layer detector type and other details required for the track building. Additionally, two-dimensional arrays of radiation and interaction lengths indexed in $r$ and $z$ are defined to account for the effect of multiple scattering and energy loss as the particle traverses the detector. The algorithm can handle either a constant or parameterized magnetic field.

\section{Physics performance}
Simulated $t\bar{t}$ events at PU 70 are used to validate the physics performance of the mkFit algorithm. To reconstruct the tracks, a constant 3.8 T magnetic field is assumed. The seed tracks used in these results correspond to the \textit{initialStep} tracking iteration of CMS, where the seeds are required to have four hits from distinct inner pixel layers and to be compatible with the beam spot constraint~\cite{CMS_tracking}. The same set of seeds is used for both the mkFit and CMSSW results.
For the calculation of the efficiency, fake rate, and duplicate rate, either simulated tracks or tracks reconstructed by the nominal CMSSW algorithm can be used as reference tracks. In both cases the reference tracks are required to have at least 12 hits, including the seed. The simulated tracks are required to have 4 hits matched to a seed track. A reconstructed track is considered matched to a reference track if at least 50\% of the hits are shared, excluding the hits from the seed. Only reconstructed tracks with at least 10 hits, including the seed, are considered.

\subsection{Efficiency}
\label{eff}

\begin{figure}[!htb]
  \centering
  \includegraphics[width=0.65\linewidth]{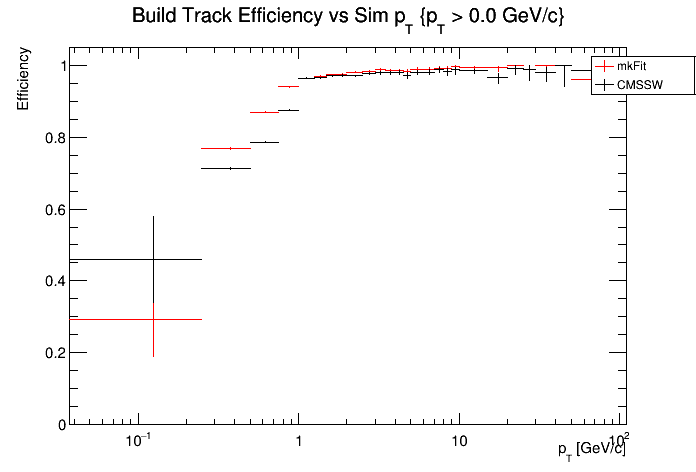}
  \caption{Efficiency of the mkFit (red) and nominal CMSSW (black) track building algorithms as a function of the track $p_T$. The efficiency is defined as the fraction of simulated tracks that are matched to at least one reconstructed track.}
  \label{fig:eff}
\end{figure}

The track building efficiency is defined as the fraction of reference tracks that are matched to at least one reconstructed track. Figure~\ref{fig:eff} shows the efficiency with respect to simulated tracks as a function of track $p_T$. The mkFit algorithm performs at least as well as CMSSW, even for low values of the track $p_T$. 

\subsection{Duplicate rate}
\label{dupremoval}

One of the primary recent accomplishments is the addition of a dedicated post-processing step to remove duplicates. The majority of the duplicates arise from duplicate seed tracks from detector module overlaps. In the nominal CMSSW tracking algorithm, the seeding region is rebuilt during the backward propagation after the initial track building. Because the tracks are processed sequentially, if all of the hits in a seed are included in a previously built track, then that seed is skipped. In the mkFit algorithm, however, this method cannot be used because many seeds are processed in parallel. The handling of duplicates in mkFit is therefore a two step process.
First, an initial seed cleaning is performed before building the tracks. The seed cleaning algorithm uses the $p_T$, $\eta$, and $\phi$ parameters of the seed tracks as well as information about shared hits to remove duplicate seeds. 
Second, after all of the tracks have been built, the track kinematic quantities ($p_T$, $\eta$, and $\phi$) and fraction of shared hits are checked and used to remove duplicates. For both steps, it is important that the requirements are not too loose, otherwise it will start to affect the efficiency for high PU events.

\begin{figure}[!htb]
  \centering
  \includegraphics[width=0.65\linewidth]{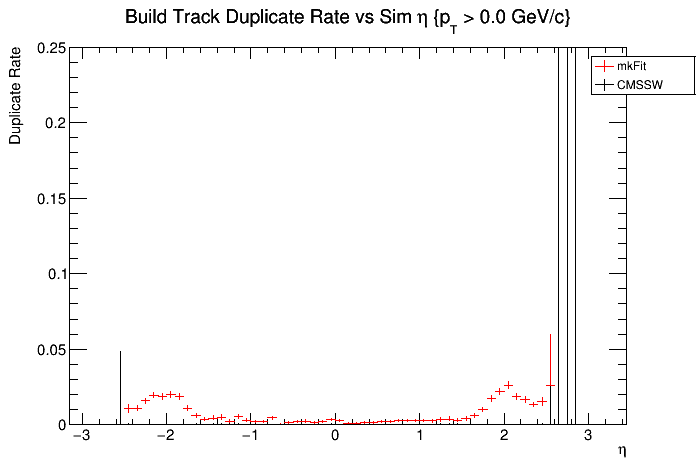}
  \caption{Duplicate rate of the mkFit (red) and nominal CMSSW (black) track building algorithms as a function of the track $\eta$. The duplicate rate is defined as the fraction of simulated tracks that are matched to multiple reconstructed tracks. The CMSSW values are nearly zero.}
  \label{fig:dr}
\end{figure}

Figure~\ref{fig:dr} shows the duplicate rate as a function of $\eta$ after the second step was implemented. The duplicate rate is defined as the fraction of reference tracks that are matched to multiple reconstructed tracks. Previously, the duplicate rate in the endcaps ($|\eta| > 1.5$) was close to 70\%. With the post-processing duplicate removal step, the duplicate rate is less than 2\% across the full range of $\eta$. Further tuning is expected to reduce the duplicate rate even more.

\subsection{Future improvements}

There is work ongoing in several areas to improve the physics performance of the mkFit algorithm. Many of the parameters used by the algorithm should be considered preliminary and will require a final optimization. This includes the size of the search windows used to find compatible hits on each layer, the scores of the candidate tracks, and the cuts used for the duplicate removal post-processing step. Additionally, the efficiency defined above only considers tracks with at least 12 hits. Dedicated efforts are needed to ensure that the high efficiency is maintained for shorter tracks as well. Finally, work is still needed to reduce the fake rate to acceptable levels.

\section{Compute performance}

\begin{figure}[!htb]
  \centering
  \includegraphics[width=0.5\linewidth]{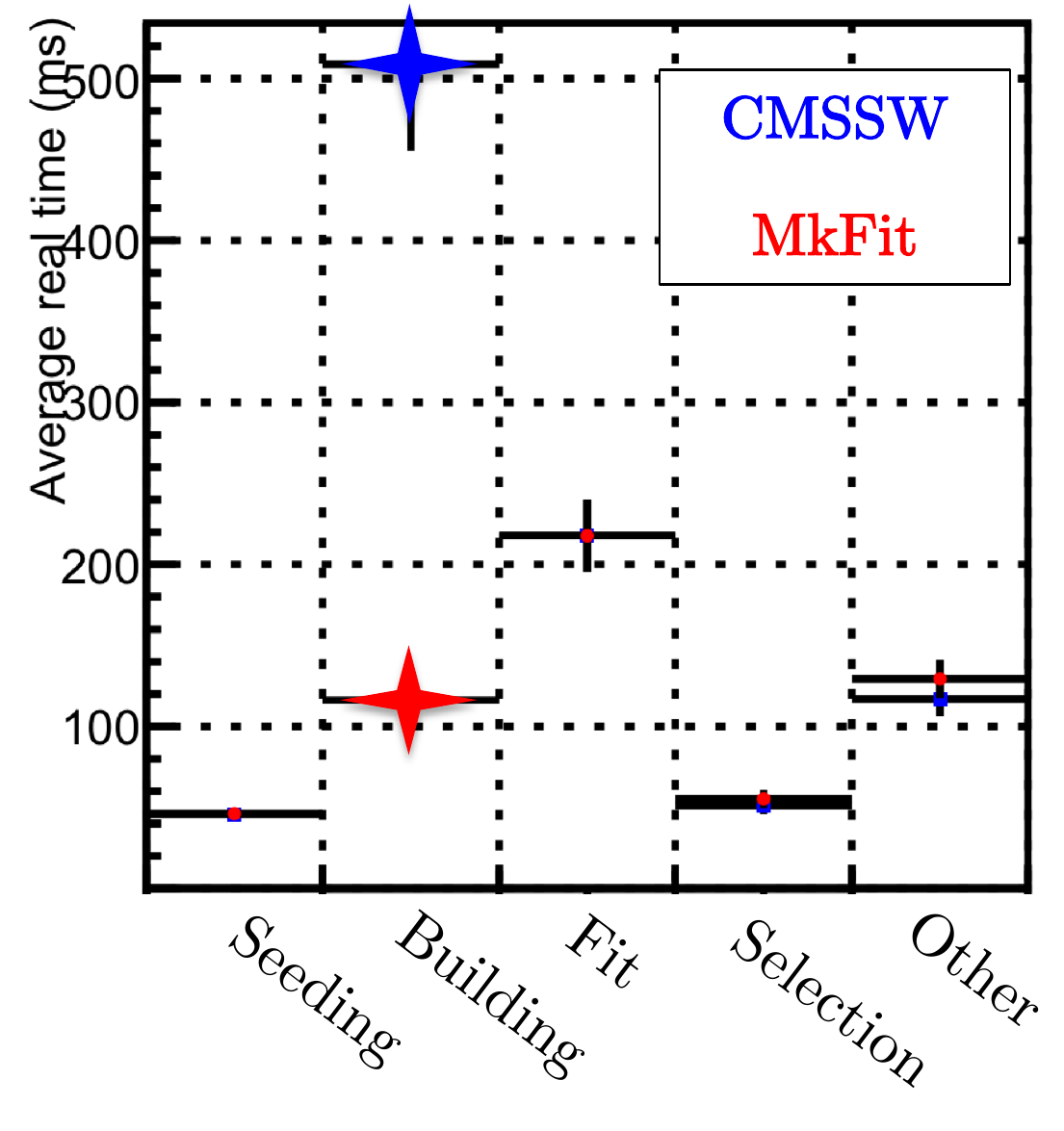}
  \caption{Average real time per event spent on each step of particle tracking when using nominal CMSSW (blue) or mkFit (red) for the building step. The other steps are all performed using the nominal CMSSW code. Results used simulated $t\bar{t}$ events with PU 50, and the code was run single-threaded on an Intel SKL-SP, Skylake Gold, Intel Xeon Gold 6130 CPU @ 2.1 GHz. The mkFit algorithm was compiled with the Intel \texttt{icc} compiler with the AVX-512 instruction set.}
  \label{fig:cmssw_time}
\end{figure}

The performance of the algorithm when run in the context of CMSSW on simulated $t\bar{t}$ events with PU 50 is shown in Figure~\ref{fig:cmssw_time}. The time is given for each part of the \textit{initialStep} tracking iteration. These results were obtained by running mkFit on a single thread on an Intel Xeon SKL-SP Gold
6130 CPU @ 2.1 GHz with 2 sockets $\times$ 16 cores with hyperthreading enabled and the Turbo Boost feature disabled. The Intel \texttt{icc} compiler was used to compile mkFit using the AVX-512 set of instructions. The mkFit algorithm is used for the ``build" step only, and the rest of the steps are performed using the normal CMSSW tracking code.

There are several notable features of this plot. First of all, the time taken in the build step is reduced by a factor of \textbf{4.3} compared to the nominal CMSSW tracking. This step includes the time spent converting the data structures into the mkFit format and back into the format expected by CMSSW. Profiling the code reveals that the data conversion takes 40\% of the total build time. If the data conversion is ignored entirely, then the mkFit algorithm is a factor of \textbf{7} times faster.

Another important thing to note is that the time taken by the track fitting step is now the same between mkFit and CMSSW. This is different from previous results shown by our group~\cite{Mario}, where the time taken for fitting was significantly longer in mkFit. The change can be attributed to the new post-processing step of the duplicate removal. Without the duplicate removal, mkFit found more duplicate tracks and the time to fit all of the tracks was correspondingly longer. 

The final important feature of Figure~\ref{fig:cmssw_time} relates to the fitting step. When using mkFit instead of CMSSW for the build step, the time for track fitting now takes \textbf{longer} than the time required for track building. This is an important achievement, as it means that track building is no longer the single most expensive step of event reconstruction.

\begin{figure}[!htb]
  \centering
  \includegraphics[width=0.65\linewidth]{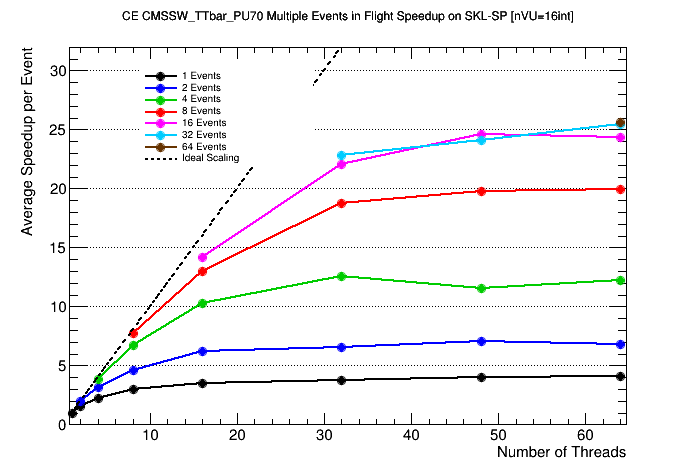}
  \caption{Speedup of the mkFit algorithm with respect to the number of threads. Different colored lines correspond to different numbers of concurrent events in flight. Results used simulated $t\bar{t}$ events with PU 70, and the code was run single-threaded on an Intel SKL-SP, Skylake Gold, Intel Xeon Gold 6130 CPU @ 2.1 GHz with the TurboBoost feature disabled. The mkFit algorithm was compiled with the Intel \texttt{icc} compiler with the AVX-512 instruction set.}
  \label{fig:MEIF}
\end{figure}

The mkFit algorithm can support multithreading and can be configured to process multiple events concurrently. The speedup as a function of the number of threads and the number of events in flight is shown in Figure~\ref{fig:MEIF} for the same Intel SKL-SP described previously. The results are measured when running mkFit in the standalone configuration on simulated $t\bar{t}$ events with PU 70. A speedup of up to a factor of approximately \textbf{25} can be achieved when processing 16 simultaneous events in flight. 

\section{Conclusions}

The performance of the mkFit algorithm continues to improve. With the addition of the new duplicate removal step, the efficiency of the algorithm is maintained and the duplicate rate is reduced to below 2\% in all regions of the detector. When running on a realistic detector geometry within the CMSSW framework, we are able to achieve a speedup of approximately a factor of 4 over the current CMSSW tracking code. This speedup can be improved to a factor of 7 if the time spent in data format conversions is minimized. Future work will include optimizing the parameters used to rank the track candidates and those used to perform the duplicate removal. The focus will be on ensuring that we are able to efficiently reconstruct short tracks ($\leq 12$ hits) as well as long tracks. The next milestone will be to incorporate the algorithm into the CMS High Level Trigger to test the algorithm online in Run 3 of the LHC. With its demonstrated physics and computational performance, the mkFit algorithm will be ready to serve as key piece of the solution to the computational challenges of the HL-LHC. 


\Acknowledgements
This work is supported by the U.S. National Science Foundation, under grants PHY–1520969,
PHY–1521042, PHY–1520942 and PHY–1624356, and under Cooperative Agreement OAC1836650, and by the U.S. Department of Energy, Office of Science, Office of Advanced Scientific
Computing Research, Scientific Discovery through Advanced Computing (SciDAC) program.


\end{document}